\documentclass{webofc}
\usepackage[varg]{txfonts}   % Web of Conferences font

\usepackage{footmisc}

\usepackage{lineno}
\def\mkfit{mkFit\xspace}

\def\stt#1{{\small\texttt{#1}}}

\def\etal{\emph{et al.}\xspace}

\def\GeVoc{\ensuremath{\,\rm{G}e\rm{V}/c}}

\def\postfigskip{\vskip-4mm}

\begin{document}

\title{Generalizing \mkfit and its Application to HL-LHC}

\author{
       \firstname{Giuseppe} \lastname{Cerati}\inst{4} %\fnsep\thanks{\email{giuseppe.cerati@cern.ch}}
  \and \firstname{Peter} \lastname{Elmer}\inst{2} %\fnsep\thanks{\email{peter.elmer@cern.ch}}
  \and \firstname{Patrick} \lastname{Gartung}\inst{4}
  \and \firstname{Leonardo} \lastname{Giannini}\inst{1}
  \and \firstname{Matti} \lastname{Kortelainen}\inst{4} %\fnsep\thanks{\email{matti.kortelainen@cern.ch}}
  \and \firstname{Vyacheslav} \lastname{Krutelyov}\inst{1} %\fnsep\thanks{\email{vyacheslav.krutelyov@cern.ch}}
  \and \firstname{Steven} \lastname{Lantz}\inst{3} %\fnsep\thanks{\email{steve.lantz@cornell.edu}}
  \and \firstname{Mario} \lastname{Masciovecchio}\inst{1} %\fnsep\thanks{\email{mario.masciovecchio@cern.ch}}
  \and \firstname{Tres} \lastname{Reid}\inst{3}
  \and \firstname{Allison} \lastname{Reinsvold Hall}\inst{5} %\fnsep\thanks{\email{ahall@fnal.gov}}
  \and \firstname{Daniel} \lastname{Riley}\inst{3} %\fnsep\thanks{\email{daniel.riley@cornell.edu}}
  \and \firstname{Matevž} \lastname{Tadel}\inst{1}\fnsep\thanks{\email{mtadel@ucsd.edu}}
  \and \firstname{Emmanouil} \lastname{Vourliotis}\inst{1}
  \and \firstname{Peter} \lastname{Wittich}\inst{3} %\fnsep\thanks{\email{wittich@cornell.edu}}
  \and \firstname{Avi} \lastname{Yagil}\inst{1} %\fnsep\thanks{\email{ayagil@physics.ucsd.edu}}
  on behalf of the CMS collaboration
}

\institute{UC San Diego, La Jolla, CA, USA 92093
  \and     Princeton University, Princeton, NJ, USA 08544
  \and     Cornell University, Ithaca, NY, USA 14853
  \and     Fermilab, Batavia, IL, USA 60510-5011
  \and     US Naval Academy, Annapolis, MD, USA 21402
}

\abstract{
\mkfit is an implementation of the Kalman filter-based track reconstruction algorithm that exploits both thread- and data-level parallelism. In the past few years the project transitioned from the R\&D phase to deployment in the Run-3 offline workflow of the CMS experiment. The CMS tracking performs a series of iterations, targeting reconstruction of tracks of increasing difficulty after removing hits associated to tracks found in previous iterations. \mkfit has been adopted for several of the tracking iterations, which contribute to the majority of reconstructed tracks. When tested in the standard conditions for production jobs, speedups in track pattern recognition are on average of the order of 3.5x for the iterations where it is used (3-7x depending on the iteration). Multiple factors contribute to the observed speedups, including vectorization and a lightweight geometry description, as well as improved memory management and single precision. Efficient vectorization is achieved with both the icc and the gcc (default in CMSSW) compilers and relies on a dedicated library for small matrix operations, Matriplex, which has recently been released in a public repository. While the \mkfit geometry description already featured levels of abstraction from the actual Phase-1 CMS tracker, several components of the implementations were still tied to that specific geometry. We have further generalized the geometry description and the configuration of the run-time parameters, in order to enable support for the Phase-2 upgraded tracker geometry for the HL-LHC and potentially other detector configurations. The implementation strategy and high-level code changes required for the HL-LHC geometry are presented. Speedups in track building from \mkfit imply that track fitting becomes a comparably time consuming step of the tracking chain. Prospects for an \mkfit implementation of the track fit are also discussed.
}

\maketitle

%%%%%%%%%%%%%%%%%%%%%%
\section{Introduction}
%%%%%%%%%%%%%%%%%%%%%%
\label{sec:intro}

The \mkfit project was started in 2014 with the goal of exploring how the traditional Kalman filter based track fitting and track finding \cite{kalmanfit} can be rethought and optimized in the age of -- at that time, novel -- many-core, vectorized computing architectures. After initial positive results on simplified detector geometries the focus was shifted to applying the \mkfit algorithm to silicon detector track finding for the CMS experiment \cite{cms-det}, resulting in a viable prototype in 2018 and culminating in a final one-year integration and validation campaign in 2021. Since the beginning of LHC Run3 in 2022 \cite{cms-run3}, \mkfit is used by CMS to reconstruct five out of twelve tracking iterations covering 90\% of found tracks with $p_T > 0.5 \GeVoc$. An in-depth review of \mkfit, including detailed motivation and algorithm description, has been published \cite{mkfit-ref}. An overview of early work with further references can be found in \cite{mkfit-chep18} and physics performance of \mkfit in CMS Run3 is available as a CMS Detector Performance note \cite{mkfit-dpnote-2022}.

This paper focuses on improvements and extensions of \mkfit that were required to support running of multiple tracking iterations in \emph{CMS software} (CMSSW) as well as to prepare it for implementation of tracking after the Phase-2 detector upgrades expected around 2030, in the \emph{High Luminosity LHC} (HL-LHC) era. Section \ref{sec:mkfit-in-cms} introduces how \mkfit is structured and run within CMSSW. A detailed presentation of required generalizations of geometry description, configuration and steering systems is given in section \ref{sec:mkit-generalization}. Currently ongoing and planned or possible future work is discussed in section \ref{sec:ongoing-future-work}.

%%%%%%%%%%%%%%%%%%%%%%%%%
\section{\mkfit in CMSSW}
%%%%%%%%%%%%%%%%%%%%%%%%%
\label{sec:mkfit-in-cms}

\mkfit was initially developed as a standalone tracking library and at first included into CMSSW as an external package. This mode of operation was used for development, physics performance tuning and benchmarking. However, one of the conditions for using \mkfit in production was for the code to be incorporated into the core CMSSW distribution, to make the software building, configuration, and patching for online and offline use compatible with CMS's requirements for computing operations. This section discusses the high-level code structure of \mkfit in CMSSW; outlines steps performed by \mkfit in a typical CMSSW reconstruction job; and, finally, presents some highlights of physics and computing performance.

\subsection{Code structure}
\label{ssec:code-structure}

\mkfit code is structured into three CMSSW packages:
\begin{itemize}
    \item \textbf{\stt{MkFitCore}} holds the central components of \mkfit, including all computational algorithms, internal data formats and geometry description, as well as configuration structures and related processing code. This core package is independent of any experiment or geometry details. It does not depend on or interact directly with any CMSSW modules or data formats.
    \item \textbf{\stt{MkFitCMS}} contains helper algorithms, called \emph{standard functions}, that perform specific tasks during track finding: seed pre-processing, candidate scoring, candidate filtering, and duplicate removal. These codes use \mkfit internal data formats and still do not depend on CMSSW.
    \item \textbf{\stt{MkFit}} is the actual bridge between \mkfit and CMSSW. It defines the CMSSW producer modules for both configuration and data-processing. It uses CMSSW specific mechanisms to pull in configuration and event-data, transforms them into \mkfit internal structures and calls appropriate steering functions. It depends both on \stt{MkFitCore} (data-formats and geometry) and \stt{MkFitCMS} (steering and standard functions) packages. 
\end{itemize}

Standalone operation of \mkfit is still possible and is frequently used for validation, tuning, development, and debugging. To support this mode, packages \stt{MkFitCore} and \stt{MkFitCMS} contain additional code and makefiles in sub-directory \stt{standalone/} that is not used by CMSSW build or touched by CMSSW code managements tools. This allows for keeping all \mkfit related files stored in a single repository. A minimal additional repository with external packages that would otherwise be used from CMSSW or CMSSW's external software still needs to be maintained separately for standalone builds.\footnote{\url{https://github.com/trackreco/mkFit-external}}

\subsection{Track finding algorithm}
\label{ssec:track-finding-algorithm}

Details of CMS track reconstruction and iterative tracking can be found in \cite{cms-tracking}. Here we are concerned with processing as it occurs for every iteration after the seed tracks have been found. Input to track finding is a vector of seed tracks, each consisting of a list of associated hits and the initial estimate of the track parameters at the final, outermost point. After that, \mkfit processing steps are as follows:
\begin{enumerate}
\setlength{\itemsep}{2pt}  \setlength{\parskip}{0pt}
    \item \emph{Seed cleaning} is performed. As \mkfit processes seeds in parallel it can not rely on hit masking in order to exclude seeds whose hits have already been consumed by previously found tracks.
    \item \emph{Seed partitioning} reshuffles the seeds into \emph{tracking regions} (barrel, transition, and end-cap). Those define the sequence in which detector layers will be visited. Additionally, the seeds are sorted in ${ \eta, \varphi }$-space to improve hit access coherency during later steps.
    \item \emph{Forward search} proceeds through the detector layers for the given tracking region going outwards from the seeding layers. For each seed, combinatorial search with a limit on the maximum concurrent number of candidates is performed, adding new hits on each layer while allowing for a limited number of missed layers and a single additional ``detector overlap'' hit. At the end, either because the edge of the tracker is reached or because no more new hits are found, the best-scoring candidate is chosen as the representative. Optionally, a quality filter can be applied before the next step.
    \item \emph{Backward fit} re-traverses each found track backwards, refitting the track parameters. If the seeding region for the current iteration does not extend all the way to the vertex region, a combinatorial \emph{backward search} can also be performed, going inwards from the first known hit.\footnote{Optionally, some or all of the seed-region hits can be dropped and new hits searched for in the seeding layers as well.} Again, the search is stopped when the innermost layers of the detector have been reached or if no new hits are found for a given seed. If the search has been performed, the best candidate is chosen as the final representative.
    \item \emph{Quality filtering \& duplicate removal} are performed on the resulting tracks.
\end{enumerate}

After the track finding for each iteration is complete, two more steps are performed by other CMSSW modules: final fit including outlier rejection; and final quality selection, based on a multivariate algorithm.

In standalone operation, the same \mkfit processing steps are performed. Pre-processed input seeds and hit data are read from a custom binary file. Final fit and quality selection are not performed, but there is a standalone version of validation comparing the found tracks to simulated ones.

\subsection{Physics and computational performance}

Here we present two highlights from the CMS Detector Performance note \cite{mkfit-dpnote-2022}.\footnote{In this note \mkfit has also been used for PixelLess iteration that was later removed due to poor performance for low-momentum highly displaced tracks.} Both cases compare relevant quantities before and after the inclusion of \mkfit in the standard Run3 track reconstruction of simulated $t\bar{t}$ events with an average pile-up of 65.

\begin{figure}[thb]
\centering
\includegraphics[width=\hsize]{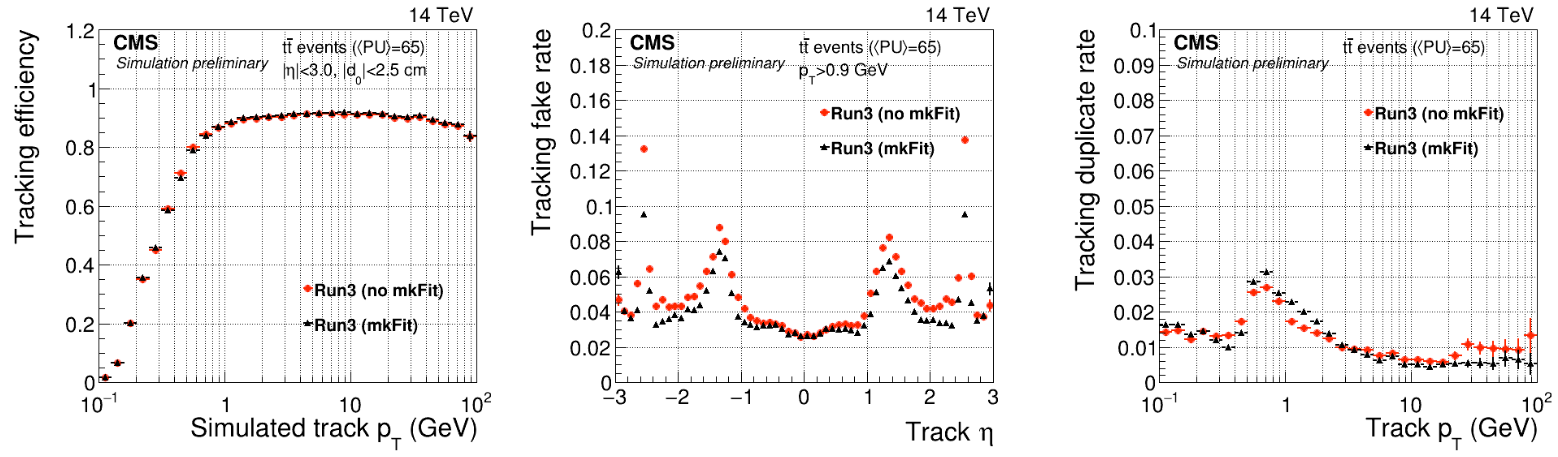}
\postfigskip
\caption{Comparison of physics performance of legacy track reconstruction (red markers) and the new Run3 tracking configuration where \mkfit is used for 6 of twelve tracking iterations (black markers): From left-to-right: a) tracking efficiency, b) fake rate, and c) duplicate rate.}
\label{fig:cms-perf}
\end{figure}

Figure \ref{fig:cms-perf} shows comparisons of basic physics performance markers. Tracking efficiency is comparable overall; efficiency vs. $\eta$ (not shown) indicates small gains in the endcap region ($2.4 < |\eta| < 2.8$). Fake rate is improved overall with reduction improving with increasing $|\eta|$. Duplicate rate is slightly increased but has been subsequently improved with further iteration-specific tuning of the duplicate removal algorithm.

\begin{figure}[thb]
\centering
\includegraphics[width=0.76\hsize]{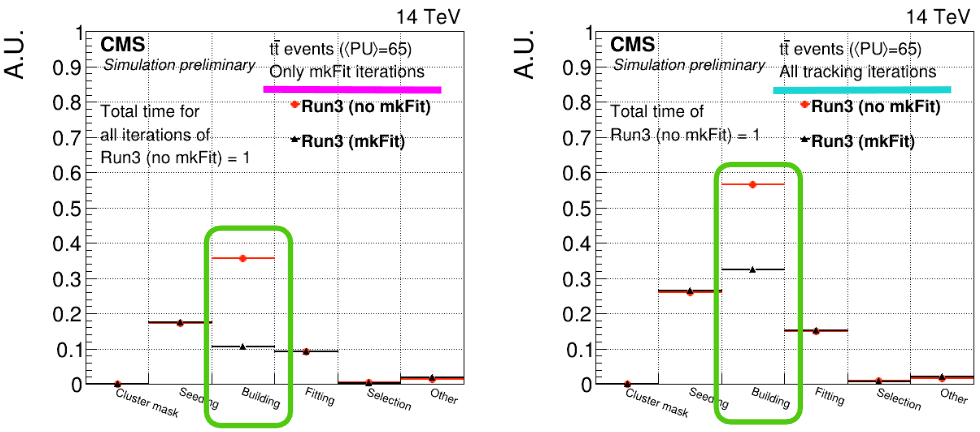}
\postfigskip
\caption{Comparison of computational performance of legacy track reconstruction (red markers) and the new Run3 tracking configuration where \mkfit is used for 6 of twelve tracking iterations (black markers). Plots show relative times of tracking steps for (left) iterations that use \mkfit and (right) all iterations.}
\label{fig:cms-speedup}
\end{figure}

Vectorization and threading scaling tests for initial iteration imply that, according to Amdahl’s Law,
\textasciitilde70\% of operations are vectorized and that more than 95\% of code is effectively parallelized. Computational speedups when using \mkfit are shown in figure \ref{fig:cms-speedup}. For all iterations where \mkfit is used, the observed track building time is reduced by \textasciitilde3.5x (the best observed reduction for one \mkfit iteration is 6.7x). Note that track building with \mkfit takes less time than seeding, and about the same time as the final fit. 

When all iterations, including non-\mkfit based ones, are considered the building time is reduced by \textasciitilde1.7x. This translates to a 25\% reduction of total tracking time (including seeding, and final fit) and, overall, results in a 10-15\% increase of Run3 reconstruction job event throughput.

%%%%%%%%%%%%%%%%%%%%%%%%%%%%%%%%%%%%%%%%%%%%%%%%%%%%%%%%%%
\section{Generalizations for iterative tracking \& HL-LHC}
%%%%%%%%%%%%%%%%%%%%%%%%%%%%%%%%%%%%%%%%%%%%%%%%%%%%%%%%%%
\label{sec:mkit-generalization}

CMSSW is a multi-threaded, module-based event processing framework that instantiates and runs modules as well as manages data sources (both event data and longer-lived data products) according to the dependencies generated by the modules themselves when the job configuration is processed. As such, each module can be instantiated multiple times and associated with different configurations and data-sources, including parallel processing of several events. This requires complete separability of configuration and module instance state as well as complete absence of any non-constant global state. Further, in the \mkfit case, each iteration has its own set of parameters that control and steer the functioning of core tracking algorithms as well as separate implementations of standard functions, as mentioned in section \ref{ssec:code-structure}.

While basic, algorithmic modifications had to be made to make \mkfit conform to the iterative tracking of CMSSW --- i.e., to support forward search, backward fit, and backward search --- the majority of these changes amounted to generalizations of the algorithms, along with mechanisms for expressing different modes of behavior through configuration structures and intermediate-level code that steers the algorithms. The previous section dealt with how the code is structured and how it operates; this section addresses the design of configuration structures and associated processing that allows the core of the code, which is independent of experiment and geometry, to run in accordance with the given detector description, algorithm tuning, and required standard functions.

\subsection{Geometry \& detector description}

In \mkfit terminology a \emph{layer} denotes an $r-z$ bounding box in global cylindrical space where hits belonging to the said layer are expected to be found. It usually corresponds in some way to detector construction or readout layers, but it does not have to:\footnote{E.g., mono and stereo hits from the same silicon-strip detector layer in CMS are split into separate \mkfit layers.} its main purposes are to aggregate the hits, provide an easy way to specify layer crossing sequences for each tracking region (called a \emph{layer plan}), and allow track search to proceed uniformly among a set of tracks. This reduces complexity and allows for the vectorization of certain key computations, including track candidate propagation, hit selection, and Kalman filter calculations.

Prior to CMSSW integration, logically dividing CMS into nested layers was sufficient to allow \mkfit to roughly reproduce the physics performance of CMSSW legacy tracking. However, as \mkfit was being considered a drop-in replacement for the existing tracking implementation, additional detector-module identification had to be included in \mkfit's geometry description to enable it to pick up multiple hits from overlapping modules within the same layer. Moreover, to support the Phase-2 upgrade geometry, which includes axially tilted detector modules, further information had to be provided (module position, normal and $\varphi$-direction vectors).

The layer boundaries, module details, and material properties are all extracted during the CMSSW job setup by traversing all inner tracker modules. For standalone usage this information gets exported into a binary file.

\subsection{Configuration structures}

As mentioned in the introduction to this section, \mkfit code needs to run concurrently within the main process, where each execution module is configured for its specific tracking iteration. As there can be no static or global data, the required configuration (or the relevant fragments) needs to be passed down the execution stack or stored in local objects. It is therefore important that the configuration data are structured in a way to facilitate such usage.

The top-level configuration for each tracking iteration is represented by the class \stt{IterationConfig}. It contains flags that control which steps of the track finding algorithm (see section \ref{ssec:track-finding-algorithm}) need to be performed, the standard functions that are to be used for this iteration (described in more detail in the next subsection), some high-level parameters for seed and duplicate cleaning, and the following structures.
\begin{itemize}
    \item Layer traversal plans for all tracking regions.
    \item Tracking parameters (e.g., maximum number of missed layers, $\chi^2$ cuts, quality filter parameters) encapsulated in class \stt{IterationParams}, with two separate instances for forward and backward search.
    \item Iteration-specific layer information, stored in class \stt{IterationLayerConfig}, which holds parameters guiding hit search and selection algorithms. These are stored in a vector, with one instance per layer.
\end{itemize}

The CMSSW module system is typically configured via Python scripts; this requires a rather tight coupling at the level of \stt{C++} code to parse incoming data. As the above \mkfit configuration is rather elaborate, it was accepted as a compromise that all \mkfit configuration can be loaded from (and saved into) JSON files. Each iteration's configuration is stored in a separate file (stored as a part of CMSSW release) and the name of this file is then passed to the \mkfit CMSSW module during instantiation.

To allow for an easy modification of a small number of parameters, reading of partial JSON overrides is fully supported: the default base configuration is read from the CMSSW release and then existing in-memory representation gets patched or overridden via simple additional JSON files or strings. Some frequently used parameters can also be set via the Python interface, e.g., to tune \mkfit performance for heavy-ion operations.

Plugin-style configuration is still supported in standalone mode and is, in fact, used to generate the default JSON files for the CMSSW operation.

\subsection{Standard function catalogs}

While adding support for multiple iterations and for Phase-2 tracking it became obvious that using a single standard function and putting additional parameters into \stt{IterationConfig} structure does not scale and that a more flexible configuration mechanism for standard functions is required for the following tasks:
\begin{itemize}
    \item seed cleaning \& partitioning – defined per iteration;
    \item candidate filters, pre- and post-backward fit – defined per iteration;
    \item duplicate cleaning – defined per iteration; and
    \item candidate scoring – defined per iteration with a possible override for each tracking region.
\end{itemize}

To provide a mechanism for registering different, specialized implementations of these standard functions, and to be able to choose them at configuration time, \emph{standard function catalogs} have been introduced. For each type of function above, a thread-safe catalog with string keys and \stt{std::function} value type is provided. The catalogs are populated via static object initializers in source files that contain the standard function codes. As \stt{std::function} objects are exported, the functions themselves can be hidden in anonymous namespaces. Further, function templates can be used to inject compile-time parameters and, in simple cases, the registered functions can be direct lambda expressions.

With this infrastructure in place, JSON files can simply specify the names (strings) associated in the catalog with the desired function. After configuration loading and setup is complete the names get resolved into \stt{std::function} objects for fast access and become available through the \stt{IterationConfig} structure.

%%%%%%%%%%%%%%%%%%%%%%%%%%%%%%%%
\section{Ongoing \& future work}
%%%%%%%%%%%%%%%%%%%%%%%%%%%%%%%%
\label{sec:ongoing-future-work}

At present, the described changes are being used to further tune Phase-1 CMS iterations. Algorithmic improvements in the processing of layers and multi-layer scoring are being investigated, with the goal of extending the usage of \mkfit beyond the current five tracking iterations, as well as improving computational performance for currently supported use-cases. In parallel, Phase-2 tracking is being developed, currently still focusing on a single, initial tracking iteration, while the specifics of track propagation and Kalman updates required for the support of tilted modules are being worked out.

The final fit is now the most time-consuming tracking task in iterations using \mkfit. With the latest additions to geometry description, it should be feasible to effectively use \mkfit for this task as well, and we are investigating the required developments in this area, along with possible improvements to existing backward-fit and backward-search algorithms.

There has been a recent intense development of CMS Phase-2 Line Segment Tracking (LST) \cite{lst-for-hllhc,lst-cms-dpnote}, a highly parallelizable algorithm that can run efficiently on GPUs, which is showing great promise for both offline and high-level trigger usage. We are planning to explore possible synergetic development with the LST project, aiming for a hybrid approach where LST performs the initial, fast track finding and \mkfit provides final steps such as backward fit, overlap hit search, and final fitting.

%%%%%%%%%%%%%%%%%%%%
\section{Conclusion}
%%%%%%%%%%%%%%%%%%%%
\label{sec:conclusion}

\mkfit is in production mode for the CMS experiment since Run3 of the LHC, used as a drop-in replacement for the legacy tracking code for five out of twelve iterations, with equivalent physics performance and with overall tracking time reduction of \textasciitilde25\%. Work has started to support CMS Phase-2 tracking geometry and events with increased pileup, where some of required changes are described in this paper, namely: generalizations of geometry description, multi-iteration configuration, and introduction of catalogs of standard functions. Exploration of extending \mkfit to also cover the final-fit and to operate synergistically with other track finding algorithms is in progress.

%%%%%%%%%%%%%%%%%%%%%%%%%%%%%%
\subsection*{Acknowledgements}
%%%%%%%%%%%%%%%%%%%%%%%%%%%%%%
\begin{acknowledgement}
This work was supported by the U.S. National Science Foundation under Cooperative
Agreements OAC-1836650 and PHY-2121686 and grant NSF-PHY-1912813.
\end{acknowledgement}

\end{document}